\documentclass[fleqn,10pt]{wlscirep}

\usepackage[utf8]{inputenc}
\usepackage[T1]{fontenc}

\usepackage{hyperref}
\hypersetup{
    colorlinks=true,
    linkcolor=blue,
    filecolor=magenta,      
    urlcolor=cyan,
    breaklinks=true,
    pdftitle={Overleaf Example},
    pdfpagemode=FullScreen,
    }

\title{FAIR for AI: An interdisciplinary and international community building perspective}

\author[1,2,*]{E.~A. Huerta}
\author[1,3]{Ben Blaiszik}
\author[4]{L.~Catherine Brinson}
\author[5,6,7]{Kristofer E. Bouchard}
\author[8]{Daniel Diaz}
\author[9,10]{Caterina Doglioni}
\author[8]{Javier~M. Duarte}
\author[11]{Murali Emani}
\author[1,2]{Ian Foster}
\author[12]{Geoffrey Fox}
\author[13]{Philip Harris}
\author[14]{Lukas Heinrich}
\author[15,16]{Shantenu Jha}
\author[17,18,19,20]{Daniel S. Katz}
\author[17,18,19]{Volodymyr Kindratenko}
\author[21]{Christine R. Kirkpatrick}
\author[22]{Kati Lassila-Perini}
\author[1]{Ravi K. Madduri}
\author[17,19,23]{Mark~S. Neubauer}
\author[24]{Fotis E. Psomopoulos}
\author[17,23]{Avik Roy}
\author[5]{Oliver R\"ubel}
\author[17,19]{Zhizhen Zhao}
\author[18]{Ruike Zhu}
\affil[1]{Data Science and Learning Division, Argonne National Laboratory, Lemont, Illinois 60439, USA}
\affil[2]{Department of Computer Science, University of Chicago, Chicago, Illinois 60637, USA}
\affil[3]{Globus, University of Chicago, Chicago, Illinois 60637, USA}
\affil[4]{Department of Mechanical Engineering and Materials Science, Duke University, Durham, North Carolina 27708, USA}
\affil[5]{Scientific Data Division, Lawrence Berkeley National Laboratory, Berkeley, CA, 94720, USA}
\affil[6]{Biological Systems \& Engineering, Lawrence Berkeley National Laboratory, Berkeley, California, 94720, USA}
\affil[7]{Helen Wills Neuroscience Institute, University of California Berkeley, Berkeley, California, 94720, USA}
\affil[8]{Department of Physics, University of California San Diego, La Jolla, California 92093, USA}
\affil[9]{Lund University, Department of Physics, Box 118, 221 00 Lund, Sweden}
\affil[10]{School of Physics \& Astronomy, The University of Manchester, Manchester M13 9PL, UK}
\affil[11]{Leadership Computing Facility, Argonne National Laboratory, Lemont, Illinois 60439, USA}
\affil[12]{Biocomplexity Institute and Department of Computer Science, University of Virginia, Charlottesville, Virginia 22904, USA}
\affil[13]{Department of Physics, Massachusetts Institute of Technology, Cambridge, Massachusetts 02139, USA}
\affil[14]{Technical University Munich, Arcisstraße 21 80333 München, Germany}
\affil[15]{Computational Science Initiative Brookhaven National Laboratory Upton, New York 11973, USA}
\affil[16]{Electrical and Computer Engineering, Rutgers, The State University of New Jersey, Piscataway, New Jersey 08854}
\affil[17]{National Center for Supercomputing Applications, University of Illinois, Urbana-Champaign, Urbana, Illinois 61801, USA}
\affil[18]{Department of Computer Science, University of Illinois at Urbana-Champaign, Urbana, Illinois 61801, USA}
\affil[19]{Department of Electrical \& Computer Engineering, University of Illinois at Urbana-Champaign, Urbana, Illinois 61801, USA}
\affil[20]{School of Information Sciences, University of Illinois at Urbana-Champaign, Urbana, Illinois 61801, USA}
\affil[21]{San Diego Supercomputer Center, University of California San Diego, La Jolla, California 92093, USA}
\affil[22]{Helsinki Institute of Physics, P.O. Box 64, 00014 University of Helsinki, Finland}
\affil[23]{Department of Physics, University of Illinois at Urbana-Champaign, Urbana, Illinois 61801, USA}
\affil[24]{Institute of Applied Biosciences, Centre for Research and Technology Hellas, Thessaloniki 57001, Greece}

\affil[*]{elihu@anl.gov}


\begin{abstract}
A foundational set of findable, accessible, interoperable, and 
reusable (FAIR) principles were proposed in 2016 as prerequisites 
for proper data management and stewardship, with the goal of enabling 
the reusability of scholarly data. The principles were also meant 
to apply to other digital assets, at a high level, and over time, 
the FAIR guiding principles have been re-interpreted or extended 
to include the software, tools, algorithms, and workflows that 
produce data. FAIR principles are now being adapted in the context 
of AI models and datasets. Here, we present the perspectives, vision, 
and experiences of researchers from different countries, disciplines, 
and backgrounds who are leading the definition and adoption of 
FAIR principles in their communities of practice, and discuss 
outcomes that may result from pursuing and incentivizing FAIR AI 
research. The material for this report builds on the FAIR for 
AI Workshop held at Argonne National Laboratory on June 7, 2022.
\end{abstract}

\begin{document}

\flushbottom
\maketitle

\thispagestyle{empty}

\section*{Introduction}

The production, collection, and curation of data require painstaking planning and the use of sophisticated experimental and computational facilities. In order to maximize the impact of these investments and create best practices that lead to scientific discovery and innovation, a diverse set of stakeholders defined a set of findable, accessible, interoperable, and reusable (FAIR) principles in 2016~\cite{fairguiding,fairmetrics}. The original intent was that these principles would apply seamlessly to data and all scholarly digital objects, including research software~\cite{FAIR4RS}, workflows~\cite{fair-workflows}, and even domain-specific custom digital objects~\cite{neubauer2022making}. However, because they were specifically written in the context of data, it became clear over time that the original set of FAIR principles would have to be translated or reinterpreted for digital assets beyond data~\cite{2022Sci377256B,2022arXiv220100247C}. This realization has led to initiatives that have proposed and/or developed practical FAIR definitions for research software and workflows, and more recently, for artificial intelligence (AI) models~\cite{2022arXiv220700611R,2022arXiv221205081D}.

In this document, we provide an inclusive and diverse perspective of FAIR initiatives in Europe and the US through the lens of researchers that are leading the definition, implementation, and adoption of FAIR principles in a variety of disciplines. This community was brought together at the \href{https://indico.cern.ch/event/1152431/}{FAIR for AI Workshop} (\url{https://indico.cern.ch/event/1152431/}) at Argonne National Laboratory on June 7, 2022. We believe that this document provides a factual, straightforward, and inspiring description of what FAIR initiatives have accomplished, what is being done and planned at the time of writing this document, and describes the end goals of these disparate initiatives. Most importantly, we hope that the ideas presented in this document serve as a motivator to reach convergence on what FAIR means, in practice, for AI research and innovation. 

\section*{FAIR Initiatives}

We have identified the following non-exhaustive list of FAIR initiatives: 

\begin{itemize}[nosep]
    \item \href{https://fair4hep.github.io}{FAIR4HEP: Findable, Accessible, Interoperable, and Reusable Frameworks for Physics-Inspired Artificial Intelligence in High Energy Physics} (\url{https://fair4hep.github.io}). 
    Funded by the US Department of Energy (DOE). In this project an interdisciplinary team of physicists, computer, and AI scientists use high energy physics as the science driver to develop a FAIR framework that advances our understanding of AI, provides new insights to apply AI techniques, and provides an environment where novel approaches to AI can be explored. 
    \item \href{https://sites.google.com/lbl.gov/endurable/home}{ENDURABLE: B\textit{en}chmark \textit{D}atasets and AI models with q\textit{u}e\textit{r}y\textit{able} metadata}  \linebreak 
    (\url{https://sites.google.com/lbl.gov/endurable/home}). Funded by DOE. The 
    goal of this project is to provide the scientific and machine 
    learning (ML) communities with robust, scalable, and extensible tools to share and rigorously aggregate diverse scientific data sets for training state-of-the-art ML models.
    \item \href{https://commonfund.nih.gov/dataecosystem}{The Common Fund Data Ecosystem} ( \url{https://commonfund.nih.gov/dataecosystem}). Funded by the US National Institutes of Health (NIH). An \href{https://app.nih-cfde.org}{online 
    discovery platform} (\url{https://app.nih-cfde.org}) that enables researchers to create and search across FAIR datasets to ask scientific and clinical questions from a single access point.
    \item \href{https://biodatacatalyst.nhlbi.nih.gov}{BioDataCatalyst} (\url{https://biodatacatalyst.nhlbi.nih.gov}). Funded by NIH. Construct and enhance annotated metadata for heart, lung, and blood datasets that comply with FAIR data principles.
    \item \href{https://thegardens.ai}{Garden: A FAIR Framework for Publishing and Applying AI Models for Translational Research 
    in Science, Engineering, Education, and Industry} (\url{https://thegardens.ai}). Funded by the US National Science Foundation (NSF). This project will reduce barriers to the use of AI methods and promote the nucleation of communities around specific FAIR datasets, methods, and AI models. Model 
    Gardens will provide a repository for models where they can 
    be linked to papers, testing metrics, known model limitations, 
    and code, plus computing and data storage resources through 
    tools such as the 
    \href{https://www.dlhub.org}{Data and Learning Hub for Science}~\cite{dlhub}, \href{https://funcx.org}{funcX}~\cite{chard2020funcx} and \href{https://www.globus.org}{Globus}~\cite{chard2016globus}. 
    \item \href{https://anl-braid.github.io/braid/}{Braid: Data Flow Automation for Scalable and FAIR Science} (\url{https://anl-braid.github.io/braid/}). Funded by DOE. This project aims to enable researchers to define sets of flows that individually and collectively implement application capabilities while satisfying requirements for rapid response, high reconstruction fidelity, data enhancement, data preservation, model training, etc. 
    \item \href{https://hpc-fair.github.io/}{HPC-FAIR: A Framework Managing Data and AI Models for Analyzing and Optimizing Scientific Applications}  \linebreak (\url{https://hpc-fair.github.io/}). Funded by DOE. This multi-institutional project aims to develop a generic High Performance Computing data management framework~\cite{hpcfair, HPCFAIR2} to make both training data and AI models of scientific applications FAIR.
    \item The \href{https://sbi-fair.github.io}{FAIR Surrogate Benchmarks
    Initiative} (\url{https://sbi-fair.github.io}). Funded by DOE.
    The research develops AI surrogates and studies their key features and the
    software environment to support their use~\cite{SABATH} in simulation
    based research. They collaborate with \href{https://mlcommons.org/en/}{MLCommons} 
    (\url{https://mlcommons.org/en/}), a
    consortium including 62 companies that host the MLPerf benchmarks,
    including those for science \cite{H3Paper,NatureBench,MLCScienceWG}, and
    mirror their processes in the computational science domain. This involves
    rich metadata involving models, datasets, and the logging of their use
    with machine and power characteristics recorded, requiring multiple
    ontologies to be developed with FAIR approaches.
    
    \item \href{https://www.materialsdatafacility.org}{The Materials Data Facility (MDF)} (\url{https://www.materialsdatafacility.org}). Funded by the National Institute of Standards and Technology (NIST) and 
    the Center for Hierarchical Materials Design, the MDF~\cite{blaiszik2016materials, blaiszik2019data} 
    aims at making materials data easily publishable,  discoverable, and reusable while following and building upon the FAIR principles. To date, MDF has collected over 80 TB of materials data in nearly 1000 datasets. In particular, this effort enables publication of datasets with millions of files or datasets comprising TB of data, and seeks to automatically index the contents in ways that provide unique queryable interfaces to the datasets. Recently, these capabilities have been augmented via the \href{https://github.com/MLMI2-CSSI/foundry}{Foundry} (\url{https://github.com/MLMI2-CSSI/foundry}) to provide access to well-described ML-ready datasets with just a few lines of Python code.
    
    \item \href{https://www.nwb.org/}{Neurodata Without Borders (NWB)} (\url{https://www.nwb.org/}). Funded by the
    NIH BRAIN Initiative. NWB is an interdisciplinary project to create a FAIR data standard for neurophysiology, providing neuroscientists with a common standard to share, archive, use, and build common analysis tools for neurophysiology data. More than just a data standard, NWB 
    is at the heart of a growing software ecosystem for neurophysiology data, including data from intracellular and extracellular electrophysiology experiments, data from optical physiology experiments, and tracking and stimulus data. A growing number of neurophysiology data
    generated by NIH BRAIN Initiative research projects and others are available on the 
    \href{https://dandiarchive.org/}{DANDI} neurophysiology data archive.

    \item \href{https://www.marda-alliance.org}{Materials Research Data Alliance (MaRDA)} (\url{https://www.marda-alliance.org}). MaRDA is an organization dedicated to helping to build community to promote open, accessible and interoperable data in materials science. MaRDA has held two virtual workshops reaching ~300 attendees last year, and has helped researchers form independent working groups. In August 2022, MaRDA leadership was funded via the NSF Research Coordination Network program to significantly expand efforts to build a sustainable community around these topics, to build consensus in metadata requirements, to train next generation workforce in ML/AI for materials, to develop shared community benchmark challenges, to host convening and coordination events, and more.

    \item \href{https://www.punch4nfdi.de}{PUNCH4NFDI} (\url{https://www.punch4nfdi.de}) is 
    the German National Research Data Infrastructure consortium of particle, astro, astroparticle, 
    hadron, and nuclear physics, representing about 9,000 scientists with a PhD in Germany from universities, the Max Planck Society, the Leibniz Association, and the Helmholtz Association. The prime goal of PUNCH4NFDI is to set up a federated and FAIR science data platform, offering the infrastructures and interfaces necessary for access to and use of data and computing resources of the involved communities and beyond. 

    \item \href{https://www.projectescape.eu}{ESCAPE} (\url{https://www.projectescape.eu}) is 
    the European Science Cluster of Astronomy \& Particle Physics ESFRI Research Infrastructures, funded from the European Union's Horizon 2020 research and innovation programme. The goal of ESCAPE is to address the critical questions of open science and long term reuse of data for science and for innovation, many of the greatest European scientific facilities in physics and astronomy have combined forces to make their data and software interoperable and open, committing to make the European Science Cloud a reality. 
    ESCAPE is delivering two Science Projects to aid the prototyping the European Open Science Cloud (EOSC) within the \href{https://eoscfuture.eu}{EOSC-Future} (\url{https://eoscfuture.eu}), another Horizon 2020-funded project. These Science Projects will advance the science, the FAIR data and the software tools needed for dark matter searches 
    and multi-messenger astronomy for extreme universe phenomena such as gravitational waves. 

    \item \href{https://github.com/tilde-lab/awesome-materials-informatics}{Awesome Materials 
    Informatics} (\url{https://github.com/tilde-lab/awesome-materials-informatics}) is an interdisciplinary and community building 
    effort to assemble a list of 
    a holistic set of tools and 
    best practices for Materials Science, encompassing 
    software and products, cloud simulation platforms, 
    and standardization initiatives.
\end{itemize}

These initiatives suggest that researchers are developing methods, approaches, and tools from scratch to address specific needs in their communities of practice. Thus, it is timely and important to identify common needs and gaps in disparate disciplines, abstract them and then create commodity, generic tools that address similar challenges across fields. Interdisciplinary efforts of this nature may leverage work led by several research data consortia which tend to be more general, e.g., the 
\href{https://www.rd-alliance.org}{Research Data Alliance (RDA)} (\url{https://www.rd-alliance.org}{), the International Science Council's Committee on Data (\href{https://codata.org/}{CODATA}) (\url{https://codata.org/}), and \href{https://www.go-fair.org/}{GO FAIR} (\url{https://www.go-fair.org/}). This translational 
approach has been showcased in the context of scientific datasets~\cite{fair_hbb_dataset} and for AI models and datasets~\cite{2022arXiv220700611R}. These recent efforts pose an important question: what is the optimal composition of interdisciplinary teams that may work together to create sufficiently generic solutions that may then be specialized down to specific disciplines and projects? As these interdisciplinary teams are assembled, and they work to define, implement, and then showcase how to adopt FAIR principles, it is critical to keep in mind that FAIR is not the goal \textit{per se}, rather the science and innovation that such principles and best practices will enable. As  well, FAIR is not a goal as much as a continual process.

In high energy physics (HEP), the experiments at the Large Hadron 
Collider at CERN are committed to bringing their data into 
the public domain~\cite{cern-data-policy} through 
the \href{http://opendata.cern.ch/}{CERN Open Data portal} (\url{http://opendata.cern.ch/}).
The CMS experiment has led the effort and, since 2014, made 
close to 3PB of research-level data public. Their 
availability opens unprecedented opportunities to process 
samples from original HEP experiment data for different 
AI studies. While the experiment data distribution follows 
FAIR principles, they remain complex, and their 
practical reusability has required further thoughts on 
the FAIR principles concretely applicable to software 
and workflows. Furthermore, the application of FAIR principles 
to data and AI models is important for the sustainability of 
HEP science and enhancing collaborative efforts with others, 
both inside and outside of the HEP domain. Ensuring that data 
and AI models are FAIR facilitates a better understanding 
of their content and context, enabling more transparent 
provenance and reproducibility~\cite{samuel2020machine,Bailey:2022tdz}. There is a strong connection between FAIRness and interpretability, as FAIR models facilitate 
comparisons of benchmark results across 
models~\cite{katz2021working} and applications of \textit{post-hoc} explainable AI methods~\cite{neubauer2022explainable}. 
As described in Ref.~\cite{Benelli:2022sqn}, data and AI models preserved in accordance with FAIR principles can facilitate education in data science and machine learning in several ways, such as interpretability of AI models, uncertainty quantification, and ease of access of data and models for key HEP use cases. In this way, they can be reliably reused to reproduce benchmark results for both research and pedagogical purposes. For instance, the detailed analysis of FAIR and AI-readiness of the CMS $H(b\overline{b})$ dataset in Ref.~\cite{fair_hbb_dataset} has explained how the FAIR readiness of this dataset has been useful in building ML exercises for open source courses on AI for HEP~\cite{duarte-course}.

In the materials science domain, the importance of 
broad accessibility of research data on all materials 
and the transformative potential impact of FAIR data 
and use of data driven and AI approaches was recognized 
with the advent of the Materials Genome Initiative (MGI) 
in 2011~\cite{MGI2011}, and with a recently released MGI 
Strategic Plan in late 2021~\cite{MGI2021}. In the decade 
since the launch of MGI, the power of integrating data 
science with materials science has unleashed an 
explosion of productivity~\cite{2022MRSBu47379D, blaiszik_aiml_stats}. Early adopters were 
computational materials scientists who launched 
a number of accessible data portals for hard materials 
and who have begun working together across the world 
on interoperability standards~\cite{2021NatSD8217A}. 
Subsequently, significant efforts have been launched 
towards capturing FAIR experimental data and tackling 
the complexities of soft materials~\cite{Brinson2020}. 
In the last several years, \href{https://www.marda-alliance.org}{MaRDA} has developed 
and flourished with multiple workshops and working 
groups addressing issues of FAIR data and models 
across all aspects of materials science.

In the life sciences, AI is becoming increasingly popular as an efficient mechanism to extract knowledge 
and new insights from the vast amounts of data that are constantly generated. AI has the potential for 
transformative impact on the life sciences since almost half of global life 
sciences professionals are either using, or are interested in using, AI in some area of their 
work~\cite{AIinHealth}. This transition is clearly shown in the explosion of ML articles in life sciences over the 
past decade: from around 500 such publications in 2010 to approximately 14k publications in 2020, an 
exponential increase that does not show any signs of slowing down in the short-term~\cite{DOME}. 
However, AI is not a one-solution-fits-all, nor a magic wand that can address any challenge in the life sciences 
and beyond. In this context, scientists pursuing domain aware AI applications may benefit from 
defining community-backed standards, such as the DOME recommendations~\cite{DOME}, 
which were spearheaded by the \href{https://elixir-europe.org/}{ELIXIR infrastructure} 
(\url{https://elixir-europe.org/}). As scientists adopt these guidelines, and prioritize openness in all aspects 
of their work processes, FAIR AI research will streamline the creation of AI applications that are trustworthy, 
high quality, reliable, and reproducible.

\subsection*{Towards a practical definition of FAIR for AI models}

There are several efforts that aim to define, at a practical level, what FAIR means for scientific datasets and AI models. 
As a starting point, researchers have created platforms that provide, in an 
integrated and centralized manner, access to popular AI models 
and standardized datasets, e.g., the \href{https://huggingface.co}{Hugging Face} (\url{https://huggingface.co}) 
platform, and the  \href{https://www.dlhub.org}{Data and Learning Hub for Science}~\cite{dlhub}.

While these efforts are necessary and valuable, additional work is needed to leverage these AI models and datasets, and translate them for AI R\&D in scientific applications. 
This is because state-of-the-art AI models become valuable tools for scientific discovery when they encode domain knowledge, and are capable of learning complex features and patterns in experimental datasets, which differ vastly from standardized datasets (ImageNet, Google's Open Images, xView, etc.). Creating scientific AI tools requires significant investments to produce, collect and curate experimental datasets, and then incorporate domain knowledge in the design, training and optimization of AI models. Often, this requires the development and deployment of distributed training algorithms in high performance computing (HPC) platforms to reduce time-to-insight~\cite{2020arXiv200308394H,KHAN2020135628}, and the optimization of fully trained AI models for accelerated inference on HPC platforms and/or at the edge~\cite{gw_nat_ast,Chaturvedi:2022suc}. 
How can this wealth of knowledge be leveraged, extended or seamlessly used by other researchers that face similar challenges in similar or disparate disciplines?

While peer-reviewed publications continue to be the main avenue to communicate advances in AI for science, researchers increasingly recognize that articles should also be linked to data, AI models, and scientific software needed to reproduce and validate data-driven scientific discovery. 
Doing so is in line with the norm in scientific machine learning, which is characterized by open access to state-of-the-art AI models and standardized datasets. 
This is one of the central aims in the creation of FAIR datasets and AI models, namely, to share knowledge, resources, and tools following best practices to accelerate and sustain discovery and innovation. 
 
Several challenges, however, need to be addressed when researchers try to define, implement and 
adopt FAIR principles in practice. This is because there is a dearth of simple-to-follow guidelines and examples, and of lack of consistent metrics that indicate when the FAIRification of datasets and AI models has been done well or not, and how to improve. 
Furthermore, while the FAIR principles are simple to read, they can be difficult to implement, and work is needed to build consensus about what they mean in specific cases, how they can be met, and how implementation can be measured, not only for data but also for other types of digital objects, such as AI models and software. The need to integrate FAIR mechanisms throughout the research lifecycle has been noted~\cite{dempsey2022sharing}. Researchers are actively trying to address these gaps and needs in the context of datasets~\cite{fair_hbb_dataset} and AI models.

On 
the latter point, two recent 
studies~\cite{2022arXiv220700611R,2022arXiv221205081D} have 
presented practical FAIR guidelines for AI models. Common
themes in these studies encompass: 1) the need to define the realm of 
applicability of these principles in the AI R\&D cycle, 
i.e., they consider AI models that have been fully trained 
and whose FAIRness is quantified for AI-driven inference; 
2) the use of common software templates to develop and 
publish AI models, e.g., the template generator \texttt{cookiecutter4fair}~\cite{cookiecutter4fair}; 
and 3) the use of modern computing 
environments and scientific data infrastructure to 
transcend barriers in hardware architectures and 
software to speak a common AI language. To ground these ideas, 
Refs.~\cite{2022arXiv220700611R,2022arXiv221205081D}  
proposed definitions of a (FAIR) AI model, 
which we have slightly modified as follows: 
``an AI model comprises a computational graph and a 
set of parameters that can be expressed as 
scientific software that, combined 
with modern computing environments, may be used to 
extract knowledge or insights from experimental or 
synthetic datasets that 
describe processes, systems, etc. 
An AI model is Findable when a digital object identifier (DOI) 
can direct a human or machine to a digital resource 
that contains the model, its metadata, instructions 
to run the model on a data sample, and uncertainty 
quantification metrics to evaluate the soundness of 
AI predictions; it is Accessible when it and its metadata may be 
readily downloaded or invoked by humans or 
machines via standardized protocols to 
run inference on data samples; 
it is Interoprable when it 
can seamlessly interact with other models, data, software, 
and hardware architectures; and it is Reusable when it 
can be used by humans, machines and other models to 
reproduce its expected inference capabiblities, and 
provide reliable uncertainty quantification metrics when 
processing datasets that differ from those originally 
used to create it and quantify its performance''.

Furthermore, 
the work presented by Ravi \textit{et al}.,~\cite{2022arXiv220700611R} emphasizes 
the need to create computational frameworks that link 
FAIR and AI-ready datasets (produced by scientific facilities or 
large scale simulations and  either hosted at data facilities or broadcast to supercomputing centers) with 
FAIR AI models (hosted at model hubs), and that 
can leverage computing environments (e.g., supercomputers, AI-accelerator machines, 
edge computing devices, and the cloud) to automate data 
management and scientific discovery. All 
these elements may be orchestrated and steered by 
\href{https://www.globus.org}{Globus} workflows.   

\subsection*{Rationale to invest in FAIR research}

There are many compelling reasons to create and share FAIR AI models and datasets. Recent studies argue that FAIR data practices are not only part of good research practices, but will save research teams time by decreasing the need for data cleanup and preparation~\cite{mons2020invest}. It is easy to dismiss anything that sounds new as an ``unfunded mandate''. However, FAIR directly relates to many compatible initiatives and goals of most scientifically focused organizations. For instance, FAIRness 
is closely connected to, and perhaps a prerequisite of, reproducibility. It is also needed for data exploration and is closely connected to ethics issues. FAIR principles can contribute to transparency and other tenets of Open Science. 

On the other hand, Supercomputing resources (e.g., Argonne Leadership Computing Facility, Oak Ridge Leadership Computing Facility, National Center for Supercomputing Applications, Texas Advanced Computing Center, etc.,) and 
scientific  data facilities (e.g., Advanced Photon Source at Argonne, National Synchrotron Light Source II at Brookhaven National Laboratory, etc.,) produce valuable data that may only be effectively shared and reused through the adoption of practical, easy to follow FAIR principles, and the design and deployment of smart software infrastructure. In brief, FAIR is an important 
step towards an 
optimal use of taxpayer dollars, it maximizes the 
science reach of large scale scientific and 
cyberinfrastructure facilities to power automated 
AI-driven discovery. 

\subsection*{Needs and gaps in AI research that may be addressed by adopting FAIR principles}

In the article that established the FAIR Principles, it is emphasized these principles should enable 
machine actionable data~\cite{wilkinson2016fair}. This is synergistic with the rapid adoption 
and increased use of AI in research. The more data is easy to locate (\textbf{F}indable), easy to 
access (\textbf{A}), well described with good and interoperable metadata (\textbf{I}), and available 
for reuse (\textbf{R}) the easier it will be to use existing data as training or validation sets for 
AI models. Specific benefits of FAIR AI research throughout the entire discovery cycle 
include:

\begin{itemize}[nosep]
    \item Rapid discovery of data via search 
    and visualization tools, ability to download data 
    for benchmarking and meta-analyses using AI for 
    further scientific discovery.
    \item Reproducibility of papers and AI models published with them.
    \item Easy-to-follow guides for how to make data and AI models FAIR are needed, as this process can be difficult, particularly for researchers to whom it is new.
    \item Establish and promote tools and data infrastructures that accept, store, and offer FAIR and 
    AI-ready data.
    \item In biomedicine and healthcare, AI models could improve 
    generalization by exposing them to diverse, FAIR datasets. 
    \item Engagement from industry partners is vital to this effort, since they are a major force in AI innovation.
    \item Get publishers involved and committed to 
    using FAIR, both for data and for other objects such as AI models and software, as they are where research results are shared.
    \item Adopting the FAIR principles in AI research will also facilitate more effective reporting. Inadequate explanations on the main parts of AI methods not only lead to distrust of the results, but also act as blockers in transferring them to an applied context, such as the clinic and patient care.
    \item Making FAIR datasets available in HEP is crucial to obtaining benchmark performances of AI models that make AI-driven discovery possible. While a large number of models have been developed for targeted tasks like classification of jets in collider experiments~\cite{kasieczka2019machine}, their performances vary with the choice of training datasets, their preprocessing, and training conditions. Developing FAIR datasets and FAIRifying AI models with well defined hyperstructure and training conditions will allow uniform comparison of these models.
    
    \item Establishing seamless and interoperable data e-infrastructures. As these infrastructures mature, a new \emph{AI services layer} will emerge; defining the FAIR principles in advance is thus important in order to accelerate this process.

    \item Computer science and AI research on efficient generic surrogate
    architectures and methods to derive reliable surrogate performance for a
    given accuracy (i.e., towards general surrogate performance models) will
    benefit extensively from FAIR data and processes.
    
     \item One element that has been often debated in AI solutions is of fair (or unbiased) models. This issue is one of the most critical in life sciences, and especially when considering applications that have a direct consequence to human health. FAIR AI and data can facilitate the overall process of identifying potential biases in the involved process.
     
    \item Where reproducibility cannot be guaranteed, FAIR data and processes can help establish at a minimum scientific correctness.
    
\end{itemize}

\subsection*{Agreed-upon approaches/best practices 
to identify foundational connections between scientific 
(meta)datasets, AI models, and hardware}

Since this work is in its infancy, there is an urgent need to 
create incentive structures to impel researchers to invest time and 
effort to adopt FAIR principles in their research, since these 
activities will lower the barrier to adopting AI methodologies. 
Adopting FAIR best practices will bring about immediate benefits. 
For instance, FAIR AI models can be constantly reviewed and 
improved by researchers. Furthermore, software can be optimized for 
performance or expanded in functionality, rather than standing 
still and stagnant. In materials science and chemistry, 
and many other disciplines, thousands of AI models are 
published each year. Thus, it is critical to rank best AI 
models, FAIRly share them, and develop APIs to streamline their 
use within minutes or seconds. Specific initiatives to address 
these needs encompass:

\begin{itemize}[nosep]
    \item \href{https://www.gofair.us}{GOFAIRUS} (\url{https://www.gofair.us}). FAIR papers that efficiently link publications, AI models, and benchmarks to produce figures of merit that quantify performance of AI models and sanity of datasets.
    \item \href{https://mlcommons.org/en/}{MLCommons} (\url{https://mlcommons.org/en/}). A consortia that brings industry and academic partners together in a pre-competitive space to compare performance of specific tasks and datasets using different hardware architectures and 
    software/hardware combinations.
    \item \href{https://thegardens.ai}{Garden} (\url{https://thegardens.ai}). A platform for 
    publishing, discovering and resuing FAIR AI models, linked to 
    FAIR and AI-ready datasets, in 
    physics, chemistry and materials science.
    \item \href{https://commonfund.nih.gov/bridge2ai}{Bridge2AI} (\url{https://commonfund.nih.gov/bridge2ai}). FAIR principles can enable ethics inquiries in datasets, easing their use by communities of practice.
\end{itemize}

While these approaches aim to ease the adoption 
and development of AI models for scientific discovery and 
to develop methods to quantify the statistical validity, 
reliability and reproducibility of AI for inference, there are 
other lines of research that explore the interplay between 
datasets, AI models, optimization methods, hardware 
architectures, and computing approaches from training through 
to inference. It is expected that 
FAIR and AI-ready datasets may facilitate these studies. 
For instance, scientific visualization and accelerated 
computing have been combined to quantify the impact of 
multi-modal datasets to optimize the performance of AI models for 
healthcare~\cite{arjun_cardio}, cosmology~\cite{asad:2018K,viztsne2}, high energy physics~\cite{2022arXiv221205081D,2022arXiv221112770R}, and observational astronomy~\cite{2020MNRAS.493.3178W,2021MNRAS.506.5294W}, 
to mention a few exemplars. These studies shed new light into the 
features and patterns that AI extracts from data to make 
reliable predictions. Similarly, recent studies~\cite{2022arXiv220312634R,pmlr-v70-kansky17a,2023arXiv230208332R} 
have demonstrated that incorporating domain knowledge in the 
architecture of AI models, and optimization methods (through 
geometric deep learning and domain aware loss functions) 
leads to faster (even zero shot) learning and convergence, and 
optimal performance with smaller training and validation 
datasets.

It is also worth mentioning that 
publishing a FAIR AI model including all relevant (meta)data, 
e.g., set of initial weights for training, 
all relevant hyperparameters, libraries, dependencies, and 
the software needed for training and optimization may not suffice 
to attain full reproducibility. This is 
because users may use 
different hardware to train and optimize AI models, and thus 
the selection of batchsize and learning rate may have to be 
adjusted if only one or many GPUs are used for distributed 
training. It may also be 
the case that users prefer to use AI-accelerator machines and 
the AI model, hyperparameters, libraries and dependencies will 
have to be changed. These considerations have persuaded 
researchers to define FAIRness in the context of AI inference. 
These caveats were also discussed by Ravi et al.~\cite{2022arXiv220700611R}, 
where a FAIR AI model was produced using distributed computing 
with GPUs, quantized with NVIDIA TensorRT, and trained from the 
ground up using the SambaNova DataScale\textsuperscript{\textregistered} system  
at the \href{https://www.alcf.anl.gov/alcf-ai-testbed}{ALCF AI Testbed} (\url{https://www.alcf.anl.gov/alcf-ai-testbed}). However, FAIRness of these different AI models 
was quantified at the inference stage.

\subsection*{Promise or role of privacy preserving and federated learning in the creation of FAIR AI datasets and models}

Sample case: \href{https://www.palisadex.net}{PALISADE-X Project} (\url{https://www.palisadex.net}). 
The scope of applications in this project includes development of AI models using closed source/sensitive data and leveraging distributed 
secure enclaves. Current applications include biomedical data, but may 
be applicable to data from smart grids, national security, physics, 
astronomy, etc. 

The development of FAIR AI tools for privacy preserving federated 
learning should be guided by several considerations. For 
instance, ethically sourced data (beyond human safety protection) 
should include attributes for the creation of AI models in a 
responsible manner. Furthermore, open, AI-driven discovery with 
protected data should be guided with clear principles and examples 
that demonstrate how to use data in a way that protects the privacy 
of individuals or organization. Ethical data sharing and 
automated AI-inference results should be regulated with input 
from interdisciplinary teams. Care should be taken to perform a 
thorough external validation of developed models to capture 
diversity and measure their applicability across different 
data distributions. In the case of personalized medicine, existing smart watches 
can identify markers that may identify suicidal behaviour. Should 
these results be readily shared with healthcare provider without 
input from individuals? These considerations demand thoughtful 
policy development and governance for datasets and AI models.

Ethical issues go well beyond biology, genomics and healthcare. 
For instance, in materials science and chemistry a recent article 
described a methodology to train an AI model to minimize drug toxicity, 
and then used to show potential misuse of maximizing toxicity for 
chemical weapons development~\cite{urbina2022dual}.

\subsection*{Transparent/interpretable AI models are considered critical to facilitating the adoption of AI-driven discovery. Why is (or isn't) this possible/reasonable in view of the ever increasing complexity of AI models?}

AI models have surpassed human performance 
in image classification challenges~\cite{NIPS2012_c399862d,chollet_image}. These algorithms process data, 
identify patterns and features in different ways to humans. 
When we try to understand what these AI models learn and how 
they make decisions, we should avoid using human-centric 
judgements 
on what is correct or acceptable. These algorithms need not work 
or ``think'' as humans to be promoted as reliable and trustworthy 
tools for scientific discovery and innovation. Rather, we 
should focus on 
defining clear, easy to follow, quantifiable principles to 
thoroughly examine AI predictions. At the same time, 
it is important to distinguish persuasive~\cite{xAI_overview}
from interpretable AI~\cite{AIRS_PB}.

Scientific visualization is a powerful tool to explore and 
get new insights on how and what AI models learn; the 
interplay among data, a model's architecture, training and  
optimization schemes (when they incorporates domain knowledge) and 
hardware used; and what triggers a sharp response in an AI model 
that is related to new phenomena or unusual noise anomalies~\cite{asad:2018K,viztsne2}.

Explainability of AI models can be deemed crucial in scientific 
domains when the decision making process of deep learning models 
can be important to make them trustworthy and generalizable. 
Interpretability of deep neural networks is important to identify 
relative importance of features and identify information pathways 
within the network. With prohibitively large complexities of 
neural architectures, existing methods of explainable AI can 
be constrained by their lack of scalability and robustness. 
Domain specific approaches for developing novel methods in explainable AI need to be explored to ensure development of reliable and reusable 
AI models ~\cite{neubauer2022explainable, khot2022detailed}.

A number of strategies to create explainable AI models include 
the use and adoption of community-backed standards for effective 
reporting of AI applications. AI practitioners should also 
define use space of a model, and evaluate resource credibility 
using, e.g., these Ten Simple Rules~\cite{erdemir2020credible}. It is also good 
practice to use well-known metrics to quantify the performance, 
reliability, reproducibility, and statistical soundness of AI 
predictions.

Current trends in explainable AI include the integration 
of domain knowledge in the design of AI architectures, training and optimization schemes, while also leaving room for serendipitous 
discovery~\cite{Stanev2021ArtificialIF,nat_state_matter}. At the end of the day we expect AI to shed light 
on novel features and patterns hidden in experimental datasets 
that current theories or phenomenology have not been able to 
predict or elucidate~\cite{ai_math}. Exploring foundation AI models, such as GPT-4~\cite{2020arXiv200514165B}, 
provides new insights on what the model has learned, and 
helps understand concepts such as model memorization 
and deep generalization.

\subsection*{Holy grail of FAIR science}

We identified the following objectives and end-goals of FAIR initiatives. 

\begin{itemize}[nosep]
    \item As stated before, FAIR is not the end-goal. It is a journey of improving practices and adapting research resources along with technology innovations. FAIR contributes by enabling discovery and innovation. It will also help us identify best practices that lead to sustainability, lasting impact, and funding. 
    \item Software, datasets, and AI models are all first class research objects. Investments and participation in FAIR activities should be considered for career advancement, tenure decisions, etc.
    \item Since digital assets cannot be open source forever (indefinite funding), FAIR initiatives should also inform what data, AI models and other digital assets should be preserved permanently.
    \item Leverage scientific data infrastructure to automate~\cite{osti_1506555} the validation and assessment of the novelty and soundness of new AI results published in peer-reviewed publications. 
    \item Create user friendly platforms that link articles with AI models, data, and scientific software to 
    quantify the FAIRness of AI models, e.g., the  \href{https://journal.physiomeproject.org/}{Physiome Project} (\url{https://journal.physiomeproject.org/}), the \href{https://reproduciblebiomodels.org}{Center for Reproducible Biomedical Modeling} (\url{https://reproduciblebiomodels.org}), and the \href{https://thegardens.ai}{Garden project}.
    \item  Recent approaches have showcased how to combine data facilities, computing resources, FAIR AI models, and FAIR and AI-ready data to enable automated, AI-driven discovery~\cite{2022arXiv220700611R}.
\end{itemize}

Creating FAIR discovery platforms for specific disciplines can possibly lead to silos, which would cut short the expected impact of FAIR initiatives. Therefore, synergies among ongoing efforts are critical to link AI model repositories, data facilities, and computing resources. This approach will empower researchers to explore and select available data and AI models. Following clear guidelines to publish and share these digital assets will facilitate the ranking of AI models according to their performance, ease of use and reproducibility; and for datasets according to their readiness for AI R\&D and compatibility with modern computing environments. This approach is at the heart of the 
\href{https://thegardens.ai}{Garden Project}, which will deliver a platform in which FAIR AI models  
for materials science, physics, and chemistry are linked to 
FAIR data, and published in a format that streamlines their use on 
the cloud, supercomputing platforms or personal computers. 
AI Model Gardens will enable researchers to cross-pollinate 
novel methods and approaches utilized in seemingly disconnected 
disciplines to tackle similar challenges, 
such as classification, regression, denoising, forecasting, etc.
As these approaches mature, and researchers 
adopt FAIR principles to produce AI-ready datasets, it will 
be possible to identify general purpose AI models, paving the 
way for the creation of \textit{foundation AI models}, which 
are trained with broad datasets and may then be 
used for many downstream applications with 
relative ease~\cite{Bommasani2021OnTO,PaLMAI,GPT4}. An 
exemplar of this 
approach in the context of materials science was presented by Hatakeyama-Sato and Oyaizu~\cite{mt_ms_AI}, in which an AI model was 
trained with diverse sources of information, including text, 
chemical structures, and more than 40 material properties. 
Through multitask and multimodal learning, this AI model 
was able to predict 40 parameters simultaneously, including 
numeric properties, chemical structures, and text.

Achieving the expected outcomes of FAIR initiatives requires coordinated scientific exploration and discovery across groups, institutions, funding agencies and industry. 
The Bridge2AI program is an example that such interdisciplinary, and multi-funding agency approach is indeed possible. Well defined, targeted efforts of this nature will have a profound impact in the practice of AI in science, engineering and industry, facilitating the cross-pollination of expertise, knowledge and tools. We expect that this document sparks conversations among scientists, engineers and industry stakeholders engaged in FAIR research, and helps define, implement and adopt an agreed-upon, practical, domain-agnostic FAIR framework for AI models and datasets that guides the development of scientific data infrastructure and computing approaches that are needed to enable and sustain discovery and innovation.



\section*{Acknowledgements}

\noindent 
\textit{E.A.H.}: 
This work was supported by the FAIR Data program of the U.S. Department of Energy, Office of Science, Advanced Scientific Computing Research, under contract number DE-AC02-06CH11357. 
\textit{B.B.}: This work was supported by the National Science Foundation under NSF Award Numbers: 1931306 and 2209892.
\textit{C.K.}: This work was supported by the National Science Foundation under NSF Award Number: 1916481 ``BD Hubs: Collaborative Proposal: West: Accelerating the Big Data Innovation Ecosystem'' and NSF Award Number: 2226453 ``Disciplinary Improvements: AI Readiness, Reproducibility, and FAIR: Connecting Computing and Domain Communities Across the ML Lifecycle''.
\textit{D.S.K., V.K., M.S.N, A.R.}: This work was supported by the FAIR Data program of the U.S. Department of Energy, Office of Science, Advanced Scientific Computing Research, under contract number DE-SC0021258. 
\textit{G.F., S.J.} This work was supported by the FAIR Data program of the U.S. Department of Energy, Office of Science, Advanced Scientific Computing Research, under contract number DE-SC0021352.
\textit{C.D., L.H.}: 
The ESCAPE project has received funding from the Horizon 2020 research and innovation programme, Grant Agreement no. 824064. The EOSC-Future project has received funding from the Horizon 2020 research and innovation programme, Grant Agreement no. 101017536. 
\textit{C.D.} has received funding from the European Research Council under the European Union’s Horizon 2020 research and innovation program (grant agreement no. 101002463) and from the Swedish Research Council.
\textit{M.E.} The HPCFAIR project is  supported by the U.S. Department of Energy,
Office of Science, Advanced Scientific Computing Program
under Award Number DE-SC0021293.

\section*{Author contributions statement}
\noindent E.A.H. conceived the convergence of ideas and visions around FAIR initiatives, and their discussion among leads of these efforts, which is documented in this manuscript. All authors reviewed and contributed 
to the manuscript.

\section*{Competing interests}
The authors declare the 
following competing interests: They are funded by the 
U.S. Department of Energy and/or the National 
Science Foundation (as described in detail in 
the Acknowledgements section) to lead the definition 
and application of FAIR principles for scientific data, 
AI models, research software, and workflows. They are 
the lead developers of scientific data infrastructure 
used to enable these advances, including Globus, funcX 
(now Globus Compute), the Data and Learning Hub for 
Science (DLHub), CookieCutter4FAIR, APPFL: Open-Source 
Software Framework for Privacy-Preserving Federated 
Learning, the Garden Project, the RDA FAIR for 
Machine Learning Interest Group, FARR: FAIR in ML, 
AI Readiness, \& Reproducibility Research Coordination 
Network, and the ELIXIR Machine Learning focus group.

\end{document}